\documentclass[aps,prab,twocolumn,amsmath,amssymb,superscriptaddress]{revtex4-2}

\pdfoutput=1

\usepackage[utf8]{inputenc} 
\usepackage{url}            
\usepackage{microtype}      
\usepackage{graphicx}
\usepackage{pgf}			
\usepackage[separate-uncertainty=true]{siunitx}
\usepackage{xfrac}			
\usepackage[hidelinks]{hyperref}
\usepackage{enumitem}  		
\usepackage{booktabs}		
\usepackage{glossaries}		
\usepackage{stackengine}    
\usepackage{tikz}
\usetikzlibrary{positioning,fit,decorations.markings,calc}
\usepackage{verbatim}		
\usepackage[capitalise]{cleveref}	

\makeglossaries
\newacronym{thz}{THz}{Terahertz}
\newacronym{3d}{3D}{three-dimensional}
\newacronym{2d}{2D}{two-dimensional}
\newacronym{vna}{VNA}{Vector Network Analyzer}
\newacronym{ptfe}{PTFE}{polytetrafluoroethylene}
\newacronym{MUT}{MUT}{Material Under Test}
\newacronym{sla}{SLA}{stereolithography}
\newacronym{dlp}{DLP}{digital light processing}
\newacronym{tosm}{TOSM}{Thru-Open-Short-Match}
\newacronym{osm}{OSM}{Open-Short-Match}
\newacronym{dut}{DUT}{device under test}
\newacronym{rf}{RF}{radio frequency}
\newacronym{fdm}{FDM}{Filter diagonalization method}
\newacronym{dlw}{DLW}{dielectric-lined waveguide}
\glsdisablehyper

\begin{document}

\title{Self-calibration technique for characterization of integrated THz waveguides}

\author{M. Kellermeier}
\email{max.kellermeier@desy.de}
\affiliation{DESY, Notkestrasse 85, 22607 Hamburg, Germany}
\affiliation{Universität Hamburg, Jungiusstraße 9-11, 20355 Hamburg, Germany}
\author{F.~Lemery}
\author{K. Floettmann}
\affiliation{DESY, Notkestrasse 85, 22607 Hamburg, Germany}
\author{W. Hillert}
\affiliation{Universität Hamburg, Jungiusstraße 9-11, 20355 Hamburg, Germany}
\author{R. Aßmann}
\affiliation{DESY, Notkestrasse 85, 22607 Hamburg, Germany}

\date{\today}

\begin{abstract}
Emerging high-frequency accelerator technology in the terahertz regime is promising for the development of compact high-brightness accelerators and high resolution-power beam diagnostics. One resounding challenge when scaling to higher frequencies and to smaller structures is the proportional scaling of tolerances which can hinder the overall performance of the structure.  Consequently, characterizing these structures is essential for nominal operation.  
Here, we present a novel and simple self-calibration technique to characterize the dispersion relation of integrated hollow THz-waveguides. The developed model is verified in simulation by extracting dispersion characteristics of a standard waveguide a priori known by theory. The extracted phase velocity does not deviate from the true value by more than $ \SI{9e-5}{\percent} $.
In experiments the method demonstrates its ability to measure dispersion characteristics of non-standard waveguides embedded with their couplers with an accuracy below $ \approx \SI{0.5}{\percent} $ and precision of $ \approx \SI{0.05}{\percent} $. Equipped with dielectric lining the metallic waveguides act as slow wave structures, and the dispersion curves are compared without and with dielectric. A phase synchronous mode, suitable for transverse deflection, is found at $ \SI{275}{GHz} $.
\end{abstract}
\maketitle

\section{Introduction} 

THz-driven accelerators have recently garnered interest for their promising acceleration gradients, compact footprints, and relatively small wavelengths which support the formation of short femtosecond electron bunches~\cite{lemery_synchronous_2018,vinatier_simulations_2018,nanni_terahertz-driven_2015,zhang_segmented_2018,hibberd_acceleration_2020,zhao_femtosecond_2020,snively_femtosecond_2020} 
and high-resolution power diagnostics~\cite{lemery_transverse_2017,li_terahertz-based_2019}.
Narrowband sources are available from both laser-based approaches~\cite{lee_generation_2000,jolly_spectral_2019} and gyrotrons, providing a path toward the usage of resonant accelerating structures ~\cite{nanni_photonic-band-gap_2017}.

Alternatively, recent developments in high-power multicycle THz production~\cite{lemery_highly_2020} support the direct use of THz waveforms in non-resonant structures, e.g., a \gls{dlw}.  In \gls{dlw}s, the structure characteristics are entirely determined by the transverse dimensions and material parameters of the structure: the inner radius $a$, outer radius $b$, dielectric permittivity of the layer $\epsilon_r$ and conductivity of the metallic shell ~\cite{flesher_dielectric_1951, zou_x-band_2001,ivanyan_wakefields_2020} -- see \cref{fig:dlw}(a).

In a DLW, an efficient interaction between the driving field and charged bunch requires synchronicity between the phase ($v_{ph}$) and bunch ($v_e$) velocities.  The phase velocity is determined by the characteristic equation of the DLW and depends on the frequency of the incident wave.  In laser-based approaches, the central frequency of the produced waveforms depend on the crystal temperature ~\cite{lee_tunable_2001}, providing a way to control the phase velocity inside the structure.  However, lower temperatures are beneficial to reduce THz absorption within the lithium niobate crystal.  Therefore, with limited tunability, it is necessary to properly characterize a structure for efficient interactions.

Several well-established techniques exist for conventional \gls{rf} structures, such as the bead-pull method~\cite{maier_field_1952,caspers1985precise} for measuring field profiles of cavities, shorting planes for determining dispersion curves~\cite{wangler_measurement_1998}, and the coaxial-wire method~\cite{faltens_analog_1971,walling_transmission-line_1989} or the recently applied Goubau-line method~\cite{sangroula_measuring_2020} for impedance measurements of vacuum components. The small dimensions of mm-scale DLWs make these techniques challenging. Alternatively, optical methods have been used to measure the dispersion characteristics of a structure in the THz regime~\cite{mitrofanov_dielectric-lined_2010}.  Here a photoconductive antenna was used to sample the electric field after the structure, providing the dispersive characteristics of the structure for various mode excitations. A limitation of this technique resides in the fact that it measures the integrated effect of the structure, i.e., it is not capable of providing local information within the waveguide, and includes coupler effects.

To eliminate the coupler contribution, two waveguides of different length, but equal cross section and same coupler geometry~\cite[Sec. 7.1.2]{healy_energy_2020} can be compared. Considering the phase difference between these two structures instead of the absolute phase shift, the phase velocity is deduced. This approach requires a precise knowledge of the length difference, and high machining precision for identical cross sectional and horn geometry. The waveguides cannot be characterized independently.

In this paper, we introduce an error network model to characterize the phase shift in integrated high frequency structures ($ \sim  \SIrange{200}{300}{GHz} $). The technique relies on a movable reflecting device within the waveguide, in analogy to a non-contacting variable short. The measured phase shift is used to deduce the phase velocity. The paper is organized as follows: first, the self-calibration method is introduced. The following section demonstrates the applicability in simulations. Then, the method is applied to experimental data. Finally, the results and limits of the error network model are discussed.

\section{Reflection method with self-calibration}
\label{sec:fourTermErrorModel}

\begin{figure}[t]
	\centering
	\includegraphics[scale=0.5]{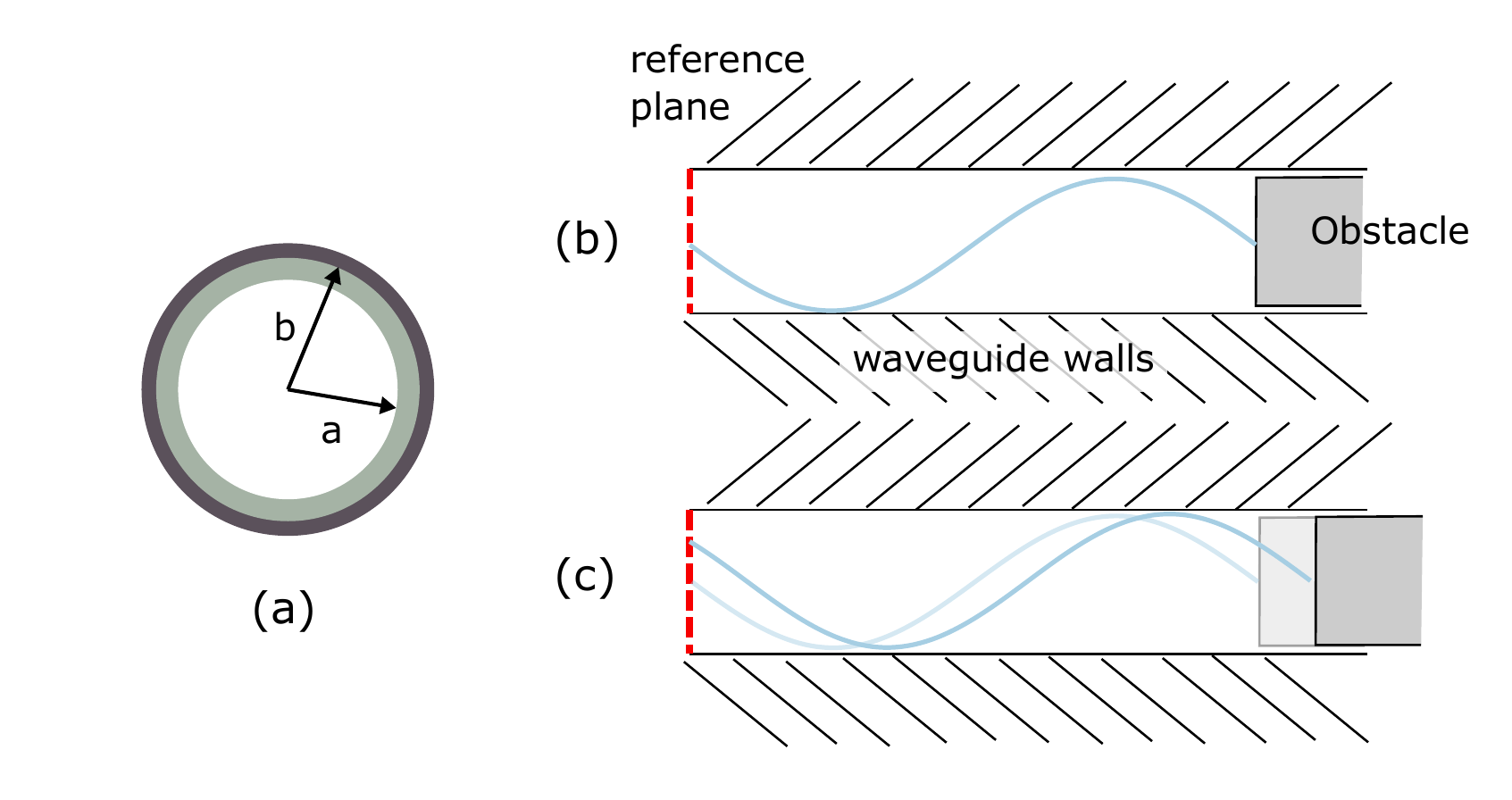}
	\caption{(a) Cross section of a dielectric loaded waveguide with a inner radius $ a $, outer radius $ b $ and permittivity $ \epsilon_r $. (b)-(c) Schematic of the reflected wave's phase shift on the reference plane when moving an obstacle through the waveguide in sub-wavelength steps.}
	\label{fig:dlw}
	\label{fig:schematic_phase_shift}
\end{figure}

In \gls{rf} network theory, scattering parameters characterize \gls{rf} devices as linear time invariant systems. The S-parameters $S_{ij}$ are complex quantities describing the ratio between outgoing traveling waves at port $ i $ when port $ j $ is excited~\cite{marks_general_1992-1}. By definition, the traveling waves are complex-valued such that S-parameters also include the phase shift. Here, it should be mentioned that the networks can also be described in terms of so called power-waves~\cite{kurokawa_power_1965}, in analogy to electrical circuits. However, the two definitions mainly differ in the renormalization of the S-matrix, while the method described here is independent of the renormalization.
In terms of \gls{rf} network theory, a waveguide is modeled as a transmission line, i.e. a 2-port device which is reciprocal, symmetric and has zero reflection coefficients if it is impedance matched to its connected devices,
\begin{equation}\label{eq:sParamsTLM}
	S= 
	\begin{bmatrix}
		0 	& e^{-\gamma L} \\
		e^{-\gamma L} & 0
	\end{bmatrix}.
\end{equation}
Here, $ \gamma = \alpha + i \beta $ is the propagation constant, including the field attenuation coefficient $ \alpha $ and the wavenumber $ \beta $, and $ L $ is the length of the transmission line. The quantity of interest here is the wavenumber as it is related to the phase velocity via $ v_{ph}= \sfrac{\omega}{\beta} $, with $ \omega $ being the angular frequency.

When short-circuiting a waveguide, the phase $ \phi_{11} := \arg S_{11} $ in reflection relates to the phase velocity via
\begin{equation}
	\phi_{11} = -2 L f \dfrac{2 \pi}{v_{ph}} \mod 2 \pi - \pi.
\end{equation}
By convention the phase is given in the interval $ [-\pi, \pi) $ instead of $ [0, 2 \pi) $.
While so-called phase-unwrapping, which adds $ 2\pi $ at jumps in $ \phi_{11}(f) $, allows for absolute phase relations across the frequency band of interest, the lowest frequency point lacks an absolute offset in multiples of $ 2\pi $. If a certain dispersion shape is already preconditioned, the offset can be determined as a free fit parameter, for instance with the analytical dispersion of a perfect electric conductor waveguide. But without prior knowledge the offset remains unknown. One way to overcome this limitation is measuring an almost identical waveguide with slightly different length $ L' $~\cite{healy_energy_2020}. Here, it is assumed that the waveguides have the same dispersion and that the length difference is on a sub-wavelength scale. Then, the phase $ \phi_{11}' $ differs by
\begin{equation}
	\phi_{11}' - \phi_{11} = -2 (L' - L) f \dfrac{2 \pi}{v_{ph}},
\end{equation}
which allows to deduce the phase velocity.

However, manufacturing imperfections could limit the similarities between both structures. In order to characterize a single waveguide independently, it requires the short circuit to be freely movable inside. As this is technically very challenging the short can be replaced by some reflecting object, denoted as an obstacle. This replacement requires that all reflections behind the object are negligible, which limits systematic errors.

\cref{fig:schematic_phase_shift}(b) and \cref{fig:schematic_phase_shift}(c) demonstrate the measurement principle. The reflecting obstacle causes a total phase shift at the input port of $ \phi = 2 l_i \beta + \phi_{\text{offset}}$, where $ l_i $ is the length of the section from the input to the reflecting device, and $ \phi_{\text{offset}} $ is an arbitrary offset caused by the reflection. When moving the obstacle by a defined step $ \Delta l $ the measured phase changes by $ \Delta \phi = 2 \Delta l \beta $, providing the unknown $ \beta$ and, thereby, the phase velocity.

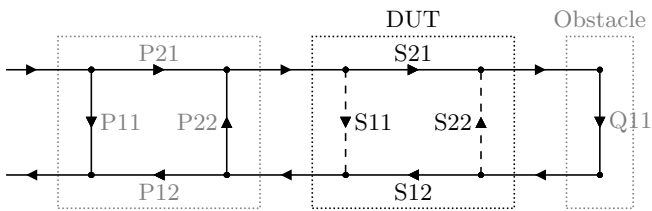
\begin{figure}[t]
	\centering
	\begin{tikzpicture}[node distance=0.7cm and 0.9cm,line width=0.6]
    \newlength{\EightPortErrorModelArrowSize}
\setlength{\EightPortErrorModelArrowSize}{0.2\baselineskip}
    \tikzset{
    	middlearrow/.style = {
    		decoration={markings,mark=at position 0.5 with {\fill (-\EightPortErrorModelArrowSize,\EightPortErrorModelArrowSize) -- (\EightPortErrorModelArrowSize,0) -- (-\EightPortErrorModelArrowSize,-\EightPortErrorModelArrowSize) -- cycle;}},
    		postaction={decorate},
    	},
	    dotStartTip/.style = {
	    	decoration={markings,mark=at position 0 with {
	    		\draw[black,fill=black] circle [radius=0.4\EightPortErrorModelArrowSize];
    		}},
	    	postaction={decorate},
	    },
	    dotEndTip/.style = {
	    	decoration={markings,mark=at position 1 with {
	    			\draw[black,fill=black] circle [radius=0.4\EightPortErrorModelArrowSize];
	    	}},
	    	postaction={decorate},
	    },
    	pics/device/.style args = {#1/#2/#3/#4/#5}{
    		code = {
    			\coordinate (-tl);
    			\coordinate [right=of -tl] (-tr);
    			\coordinate [right=of -tr] (-tr);
    			\coordinate [below=of -tr] (-r);
    			\coordinate [below=of -r] (-br);
    			\path (-tl) |- coordinate[pos=0.5] (-bl) (-br);
    			
    			\draw[middlearrow, dotStartTip, dotEndTip] (-tl) -- node[pos=0.5,above] (l21) {\color{#5}#1#3#2} (-tr);
    			\draw[middlearrow, dotStartTip, dotEndTip] (-br) -- node[pos=0.5,left] (l22) {\color{#5}#1#3#3} (-tr);
    			\draw[middlearrow, dotStartTip, dotEndTip] (-br) -- node[pos=0.5,below] (l12) {\color{#5}#1#2#3} (-bl);
    			\draw[middlearrow, dotStartTip, dotEndTip] (-tl) -- node[pos=0.5,right] (l11) {\color{#5}#1#2#2} (-bl);
    			
    			\node[draw,densely dotted,color=#5,fit=(-tl) (-tr) (-br) (-bl),inner sep=1.2\baselineskip,label={\color{#5}#4}] (-bb) {};
    		}
    	},
    	pics/transmissionline/.style args = {#1/#2/#3/#4/#5}{
    		code = {
    			\coordinate (-tl);
    			\coordinate [right=of -tl] (-tr);
    			\coordinate [right=of -tr] (-tr);
    			\coordinate [below=of -tr] (-r);
    			\coordinate [below=of -r] (-br);
    			\path (-tl) |- coordinate[pos=0.5] (-bl) (-br);
    			
    			\draw[middlearrow, dotStartTip, dotEndTip] (-tl) -- node[pos=0.5,above] (l21) {\color{#5}#1#3#2} (-tr);
    			\draw[middlearrow, dotStartTip, dotEndTip, dashed] (-br) -- node[pos=0.5,left] (l22) {\color{#5}#1#3#3} (-tr);
    			\draw[middlearrow, dotStartTip, dotEndTip] (-br) -- node[pos=0.5,below] (l12) {\color{#5}#1#2#3} (-bl);
    			\draw[middlearrow, dotStartTip, dotEndTip, dashed] (-tl) -- node[pos=0.5,right] (l11) {\color{#5}#1#2#2} (-bl);
    			
    			\node[draw,densely dotted,color=#5,fit=(-tl) (-tr) (-br) (-bl),inner sep=1.2\baselineskip,label={\color{#5}#4}] (-bb) {};
    		}
    	},
    	pics/obstacle/.style args = {#1/#2/#3/#4/#5}{
    		code = {
    			\coordinate (-tl);
    			\coordinate [right=of -tl] (-tr);
    			\coordinate [right=of -tr] (-tr);
    			\coordinate [below=of -tr] (-r);
    			\coordinate [below=of -r] (-br);
    			\path (-tl) |- coordinate[pos=0.5] (-bl) (-br);
    			
    			\draw[middlearrow, dotStartTip, dotEndTip] (-tl) -- node[pos=0.5,right] (l11) {\color{#5}#1#2#2} (-bl);
    			
    			\node[draw,densely dotted,color=#5,fit=(-tl) (-tl) (-bl) (-bl),inner sep=1.2\baselineskip,label={\color{#5}#4}] (-bb) {};
    		}
    	}
    }
    
    \pic (O) {device = P/1/2//gray};
    \pic[right=of O-tr,shift=($1.5*(O-bb.east)-1.5*(O-r)$)] (S) {transmissionline = S/1/2/DUT/black};
    \pic[right=of S-tr,shift=($1.5*(O-bb.east)-1.5*(O-r)$)] (Q) {obstacle = Q/1/1/Obstacle/gray};
    
    \draw[middlearrow] (O-tr) -- (S-tl);
    \draw[middlearrow] (S-bl) -- (O-br);
    
    \draw[middlearrow] (S-tr) -- (Q-tl);
    \draw[middlearrow] (Q-bl) -- (S-br);
    
    \path (O-tl) -| coordinate (1) ($2.5*(O-bb.west)$);
    \path (O-bl) -| coordinate (2) ($2.5*(O-bb.west)$);

	\draw[decoration={markings,mark=at position 0.3 with {\fill
			 (-\EightPortErrorModelArrowSize,\EightPortErrorModelArrowSize) -- (\EightPortErrorModelArrowSize,0) -- (-\EightPortErrorModelArrowSize,-\EightPortErrorModelArrowSize) -- cycle;}},
	    postaction={decorate}
    ] (1) -- (O-tl);
    \draw[decoration={markings,mark=at position 0.7 with {\fill
    		(-\EightPortErrorModelArrowSize,\EightPortErrorModelArrowSize) -- (\EightPortErrorModelArrowSize,0) -- (-\EightPortErrorModelArrowSize,-\EightPortErrorModelArrowSize) -- cycle;}},
    postaction={decorate}
    ] (O-bl) -- (2);
    
\end{tikzpicture}
	\caption{Error Network model, summarizing the couplers and the free space as $ P $-network and describing the obstacle as $ Q $-network. The \gls{dut} represents the waveguide which imposes $ S_{11}=S_{22}=0 $.}
	\label{fig:errorModelNtwk}
\end{figure}

However, since the waveguide is embedded with its couplers, it cannot be characterized independently. An error network model is constructed, \cref{fig:errorModelNtwk}, which summarizes all elements upstream of the embedded \gls{dut} as $ P $-network. Then, the measured reflection coefficient $ S_{11}^{(m)} $ at the test port reads as,
\begin{equation}\label{eq:s11fromNtwk}
	S_{11}^{(m)} = P_{11} + Q_{11} \dfrac{P_{21}P_{12} }{ k^{-2} - P_{22} Q_{11}} , \quad k := e^{- \gamma l},
\end{equation}
where $ P_{ij} $ and $ Q_{11} $ denote the S-parameters of the P-network and the Q-network, respectively. $ S_{ij} $ of the \gls{dut} are substituted by \cref{eq:sParamsTLM} in the derivation, but with a shorter effective length $ l, l < L ,$ of the transmission line due to the obstacle being inserted. The goal of a calibration is to extract the \gls{dut}'s S-parameters from \cref{eq:s11fromNtwk}. In fact, the network model is an extension of the standard three-term error model for the measurement of a 1-port device~\cite{rumiantsev_vna_2008, hand_developing_1970}, which is calibrated via \gls{osm} standards. The terms $ P_{11}, P_{21}P_{12} $ and $ P_{22} $ correspond to directivity $ E_{D} $, source match $ E_{S} $ and reflection tracking $ E_{R} $ in the error terms representation, respectively. The \gls{dut} from the three-term model is replaced by the reflection coefficient $ k^2 Q_{11} $. Due to the additional unknown propagation constant the model is denoted as the four-term error model.
However, since the \gls{dut} cannot be detached from the couplers, no calibration standards can be applied.

Placing the obstacle at four different positions $ l_{i} $ with steps smaller than the free space wavelength, is in principle sufficient in order to solve the system of equations
\begin{equation}\label{eq:iThs11fromNtwk}
	S_{11}^{(i)} = a + \dfrac{b}{e^{-2(\alpha + i \beta) l_i} - c}, \quad  i \in [1,2,3,4],
\end{equation}
for the four unknowns $ a, b, c $ and $ \beta $. Here, the error terms from \cref{eq:s11fromNtwk} are summarized as arbitrary unknowns $ a, b $ and $ c $.
However, it turns out to be numerically more stable to over-determine the system of equations and, instead, apply a non-linear least squares algorithm to the measured reflection coefficients $ S_{11}(l) $ depending on the obstacle position. In order to include bounds for the solution of \cref{eq:iThs11fromNtwk} the trust region reflective algorithm is chosen for the non-linear least-squares problem, as implemented in the \textit{scipy} package~\cite{2020SciPy-NMeth,branch1999subspace}.

In \cref{eq:iThs11fromNtwk}, the attenuation is constrained to be zero, $ \alpha = 0 $. For most \gls{dut}s, especially the metallic waveguides, the attenuation is too small to be significant on the length scale of the obstacle position sweep. For the case of lossy waveguides the attenuation is significant, and can be included in the fit routine, as shown in \cref{sec:unknownWvgExperiments}.

During a measurement, the \gls{vna} sweeps the excitation frequency within the band of interest, between $ \SI{220}{\giga\hertz}$ to $\SI{330}{\giga \hertz} $. This is repeated for each obstacle position such that the final data set includes both dependencies, $ S_{11}(f, l) $. The least-squares problem is solved for each frequency point $ f_{j} $ independently, which results in the phase velocity $ v_{ph}(f_{j}) = \sfrac{2 \pi f_j}{\beta_j} $.

In the next section the developed method is validated by simulations, before application to experimental data in the subsequent sections.

\section{Simulation}
\begin{figure}
	\includegraphics[width=0.4\textwidth]{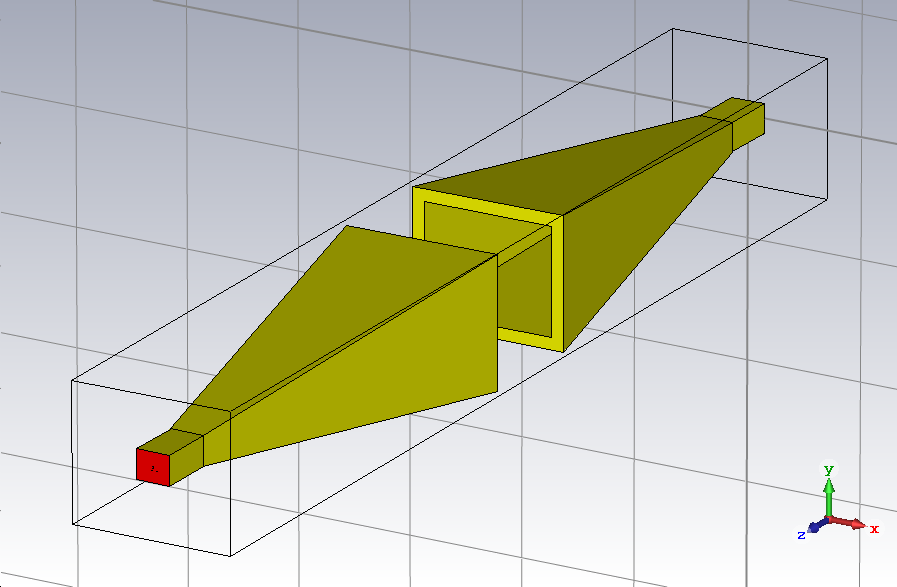}
	\caption{Out- and In-coupling horns considered as single network device with simulated scattering parameters.}
	\label{fig:horn2horn}
\end{figure}
In order to construct a network reasonably described by the error model from \cref{sec:fourTermErrorModel}
a coupler-to-coupler transition was simulated with the time domain solver of CST Studio Suite 2019~\cite{noauthor_cst_2020}.
The overall transition consists of a WR3.4 waveguide port attached to a pyramidal out-coupling horn, a free space section of $ \SI{10}{mm} $, another horn coupler of the same shape for in-coupling and the final same WR3.4 waveguide port, see \cref{fig:horn2horn}. The rectangular waveguide has a standard cross section of $ \SI{0.864}{mm} $ by $ \SI{0.432}{mm} $ while the horn has an aperture cross section of $ \SI{11.14}{mm} $ by $ \SI{8.95}{mm} $ after a transition of $ \SI{35}{mm} $ length. The dimensions are based on the optimal horn condition~\cite{balanis_pyramidal_2012}.
Here, for simplicity the out- and in-couplers were considered to be identical and the free space section is kept short for a smaller computational domain. While these two conditions are not met in the experiments in \cref{sec:experiment}, the error network is capable of modeling both cases.

\begin{figure}
	\includegraphics[width=0.43\textwidth]{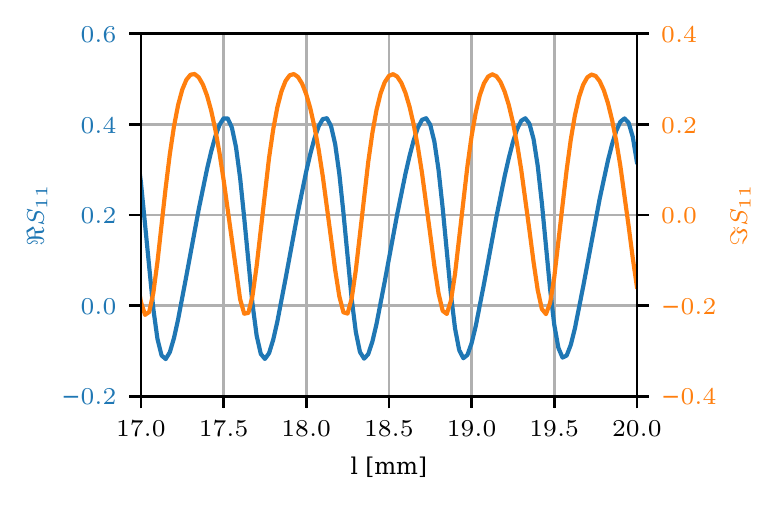}
	\caption{Real and imaginary parts of the simulated reflection coefficient $ S_{11}(l) $ as a function of the obstacle position,
		at a fixed frequency of $ \SI{303.16}{GHz} $ based on the WR3.4 waveguide as \gls{dut} and an arbitrary generated obstacle reflection. The asymmetry of $\Re S_{11} $ reveals the difference compared to a sine shape.}
	\label{fig:simulated_raw_s11}
\end{figure}
At the ports the horn-to-horn transition was excited in the frequency range of interest from $ \SI{220}{GHz} $ to $ \SI{330}{GHz} $ in order to compute the scattering parameters. These were further processed with the \gls{rf} network modeling package \textit{scikit-rf}~\cite{hillairet_rf_2020-2,arsenovic_scikit-rf-open_2018,scikit-rf_developers_scikit-rfscikit-rf_2020}. The horn-to-horn transition is treated as the $ P $-network from \cref{fig:errorModelNtwk}, while the $ Q $-network is constructed from a randomly generated S-parameter, which is constant with respect to frequency. Both networks are cascaded with a transmission line of adjustable length $ l_{i} $ in between. The transmission line is modeled with the geometrical parameters of the WR3.4 waveguide. \cref{fig:simulated_raw_s11} shows an example of the computed reflection coefficient in terms of real and imaginary parts depending on the obstacle position for a fixed frequency. While the periodicity of $ 1/2 \beta $ is obvious, one can also observe that $ S_{11}(l) $ does not follow a simple sine shape. The term $ a $ from \cref{eq:iThs11fromNtwk} leads to an offset, while $ b $ and $ c $ distort the symmetry around a peak. Especially the real part shows a flatter slope on the left side while the slope to the right is steeper.

\begin{figure}
	\includegraphics[width=0.43\textwidth]{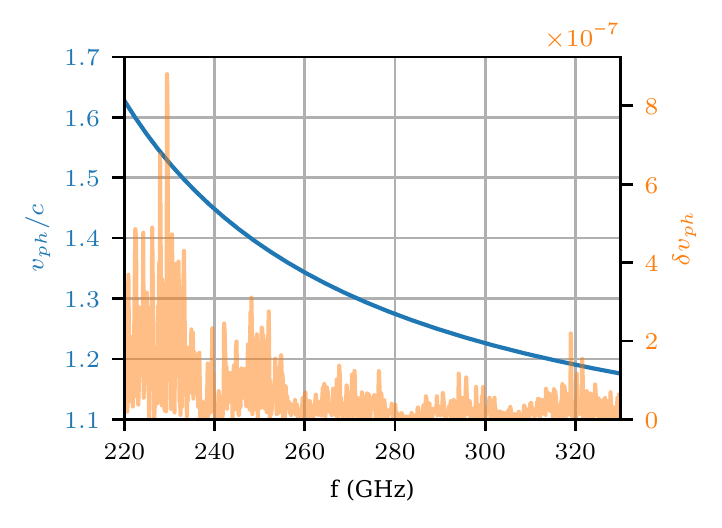}
 	\caption{Phase velocity $ v_{ph} $ (blue line) extracted from the simulation using the four-term error model and its relative deviation $ \delta v_{ph} $ (orange line) from the predicted known phase velocity for the WR3.4 waveguide.}
	\label{fig:simulated_v_ph_and_delta}
\end{figure}

The goal of fitting the four-term error model is to extract the phase velocity and verify it with the expected one. In addition, the model outputs the error terms $ a,b, c $ from \cref{eq:iThs11fromNtwk}. To find a suitable starting point, $ \beta_0$, for the parameter $ \beta $ in the least squares algorithm, a fast Fourier transform is applied to $ S_{11}(l) $ and the point of maximum amplitude in the reciprocal space is chosen. 
For the frequency point in \cref{fig:simulated_raw_s11} the fit results in a wavenumber $ \beta = \SI{5.209}{mm^{-1}} $ which gives $ v_{ph} = \SI{1.220}{c} $. In \cref{fig:simulated_v_ph_and_delta} the phase velocity is plotted over the whole frequency range considered, together with its relative deviation from the known reference $ v_{ph}^{WR3.4} $, 
\begin{equation}\label{eq:delta_v_ph}
	 \delta v_{ph} = \dfrac{ \left|v_{ph} - v_{ph}^{WR3.4}  \right |}{v_{ph}^{WR3.4}} \quad.
\end{equation}
The extracted values do not differ from the input reference values by more than $ \SI{9e-5}{\percent}$. Based on the scattering parameters of $ P $ and $ Q $ used in constructing the network, the error terms can also be compared to the prediction, where $ a = P_{11} $ is expected as well as $ b= Q_{11} P_{21} P_{12} $ and $ c =P_{22} Q_{11} $. The fit parameter $ a $ is the most precise one of the three as it describes the offset of $ S_{11}(l)$. Across the whole frequency range it deviates by $ \SI{0.04}{\percent}$ on average. At frequencies where $ P_{11} \approx 0 $ the relative deviation increases significantly to about $ \SI{3.4}{\percent}$. The other two fit parameters $ b $ and $ c $ have a similar effect on the shape of $ S_{11}(l)$ such that the fit routine can determine them less precisely. As before, their deviations from the expected value is largest when they approach zero. On average, they deviate by $ \sim \SI{7}{\percent} $.

Since the fit parameter of interest is $ \beta $, deviations up to $ \SI{10}{\percent} $ in the error terms are acceptable. With the preceding simulation, the four term error model has demonstrated its applicability in measuring the phase velocity of an integrated waveguide. Here, the WR3.4 waveguide supports only the fundamental mode which is why the horns are placed in close proximity. For unknown devices the phase front error may couple power to higher order modes if present. In the experiment this is mitigated by placing the horn's further apart from each other.

\section{Experiment}
\label{sec:experiment}
The setup consists of a Rohde \& Schwartz ZVA67 \gls{vna} which is connected to a ZC330 millimeter-wave converter spanning the frequency range between $ \SI{220}{GHz} $ and $ \SI{330}{GHz} $. The converter provides a rectangular waveguide output port of type WR3.4 to which a standard gain horn antenna is attached (Anteral, SGH-26). In order to compensate internal systematic errors the output port was calibrated via a standard \gls{osm} calibration. Errors up to the waveguide flange, such as directivity errors from the cables attached, are not covered by the above introduced error model. So calibrating up to the flange moves the reference plane for the S-parameter measurement to the connector between waveguide and horn antenna.

The monolithic \gls{dut} was mounted on an optical post in line with the beam axis of the out-coupling horn. A rail allows different positioning along the axis which helps with in-coupling. The free-space distance of $ \sim \SI{5}{cm} $ was chosen as compromise between lower phase error at further distances and higher incident power at closer distances.

The obstacle was mounted on a motorized stage on top of the rail, downstream of the structure. The choice for obstacles is restricted by being non-contacting movable inside small waveguide apertures in the order of $ \approx \SI{1}{mm} $ diameter. Thin stiff devices such as a blunt needle of a dosing syringe, and blue steel wires were applied.
The gauge 22 needle has an outer diameter of $ \SI{0.72}{mm} $ and nominal inner diameter of $ \SI{0.51}{mm} $. The set of blue steel wires covers a range from $ OD = \SI{0.3}{mm} $ to $ OD = \SI{4}{mm} $ while only wires with $ OD = \SI{0.50}{mm}, \SI{0.81}{mm}, \SI{1.21}{mm} \text{ and } \SI{1.40}{mm}$ have been used in the present study so far.
All measurements have been conducted at ambient air at $ \SI{21}{\degreeCelsius} $ and relative humidity of $ \SI{50}{\percent} $. Dry air contributes to the refractivity with $ N_{dry} = (n - 1) \times 10^{6} \approx \num{270}$~\cite[Sec. 1.3.5.1]{smith_refractive_1993}, while the frequency independent part of water vapor contributes $ N_{H_{2}O} \approx 70$~\cite{yang_time_2012}. Although water vapor provides $ \num{49} $ absorption lines in the frequency range of interest~\cite{gordon_hitran2016_2017,kochanov_hitran_2016-1},
even the highest one at $ \SI{325}{GHz} $ contributes only by about $ N_{\SI{325}{GHz}} \approx 0.3 $, based on the phase shift lineshapes~\cite{yang_understanding_2012, he_physics-based_2019}. 
 
\begin{figure*}
	\includegraphics[width=0.8\textwidth]{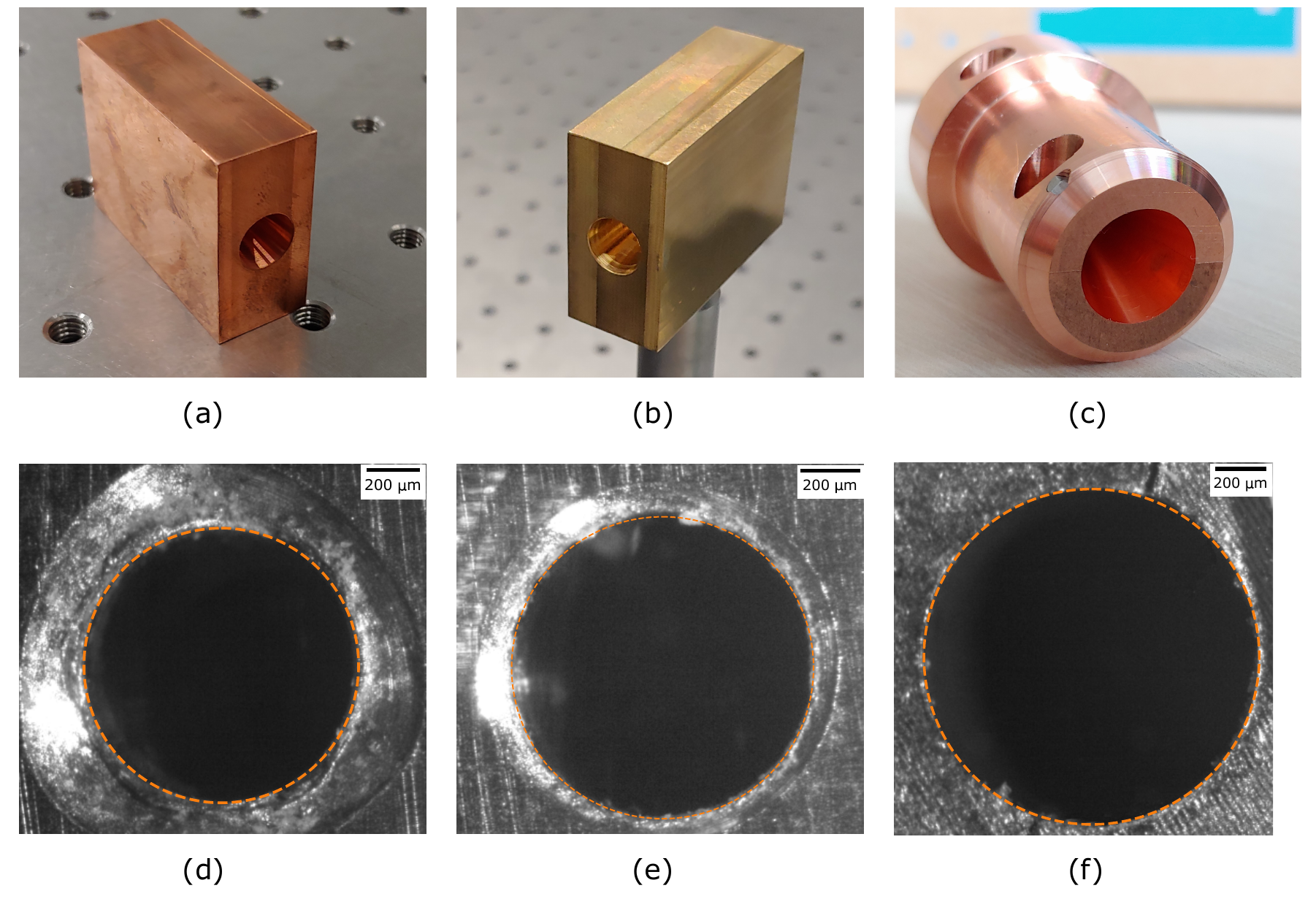}
	\caption{Metallic waveguides under test. They are not lined with dielectric. (a) Copper structure with drilled channel (b) Same design but machined in brass (c) Split-block structure fabricated by wire EDM. (d) - (f) Microscope images of the apertures with measured radii $ a $ of $ \SI{511 \pm 10}{\micro m} $, $ \SI{505 \pm 7}{\micro m} $, and $ \SI{657 \pm 7}{\micro m} $, respectively.}
	\label{fig:allWvg_imgs}
\end{figure*}
\subsection{Results for metallic waveguides}

Three circular metallic waveguides have been tested, and compared to predictions based on the known analytical dispersion relation~\cite{pozar2011microwave},
\begin{equation}\label{eq:metallicWvgDisp}
	v_{ph}/c = \dfrac{1}{n}\dfrac{1}{\sqrt{1- \left (\dfrac{f_c}{f} \right )}}, \quad 	f_c = \dfrac{c u_{11}}{2\pi a}, \quad u_{11} = \num{1.841},
\end{equation}
where $ f_{c} $ is the cutoff frequency, $ a $ the radius of the cross section, and $ u_{11} $ the first root of the derivative of the first order Bessel function, $ J_1'(u_{11}) = 0 $, due to the fundamental mode being the $ TE_{11} $ mode. Although the measurements are not conducted in vacuum, it is assumed for simplicity that the refractive index $ n = \sqrt{\epsilon_r} = 1 $. The error introduced by the assumption is in the order of $ \SI{0.03}{\percent} $, which is below the current limitations in experimental uncertainties. Microscope images of the waveguide apertures were taken and analyzed by a circle detection, using the \textit{scikit-image} package~\cite{van_der_walt_scikit-image_2014}. The algorithm performs a Hough transform~\cite{hart_use_1972} of the detected edges in the image and returns a score for each pair of pixel and radius. The pair with the highest score is identified as the center and the radius of the aperture. The peak width compared to the background of the score is considered as the uncertainty. \cref{fig:allWvg_imgs} shows the three structures, the first one being a copper block with the waveguide being drilled, the second one was also drilled but in brass, and the third one is again a copper structure which was wire electron-discharge machined (EDM). Due to machining limitations, the structure was manufactured in two half shells which were precisely assembled afterwards with alignment pins. This structure is referred to as split block waveguide.
The two different machining techniques have significantly different final features.  
A ridge formation is observable from the bored structures while the EDM provides a much more explicit edge.

\begin{figure}
	\includegraphics[width=0.43\textwidth]{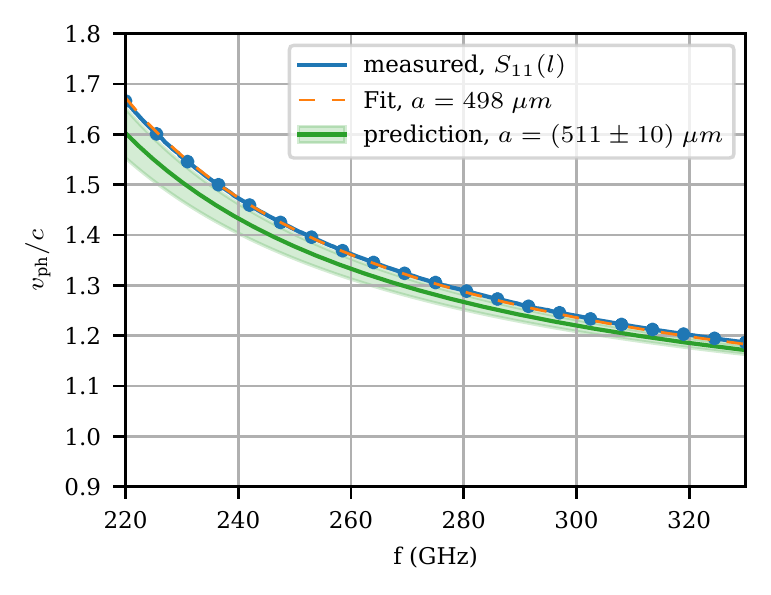}
	\caption{Measured phase velocity dispersion (blue) and fit with the analytical model (orange, dashed) of the drilled copper waveguide, shown in \cref{fig:allWvg_imgs}(a). The prediction (green) with uncertainty (green band) is based on radius measurements from \cref{fig:allWvg_imgs}(d), including errors in circle detection and finite pixel resolution. The measurement does not fall into the $ 1\sigma $-uncertainty band of the prediction. Uncertainties in the measurement are too small to be visible on the chosen axis scale. A few selected points of the measurement are highlighted as bullets for improved visualization with respect to the fit.}
	\label{fig:dispersion_wvg01}
\end{figure}

\begin{figure}
	\includegraphics[width=0.43\textwidth]{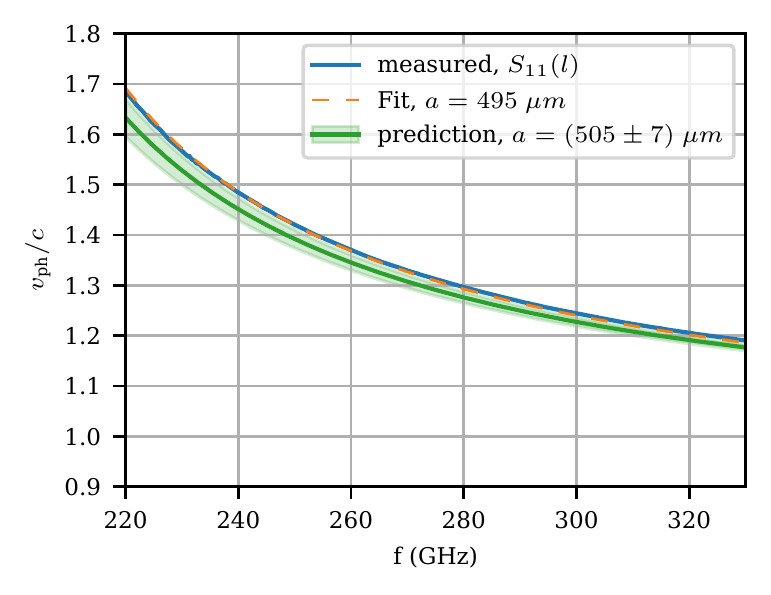}
 	\caption{Measured (blue line), fitted (orange, dashed) and predicted phase velocity dispersion of the brass waveguide, shown in \cref{fig:allWvg_imgs}(b). Prediction based on \cref{fig:allWvg_imgs}(e).}
	\label{fig:dispersions_drilled_wvgs}
	\label{fig:dispersion_wvg02}
\end{figure}

The scans of the obstacle position covered a range of $ \SI{10.00}{mm} $ inside the hollow channel, with minimum distance to the free space transition of $ \SI{5}{mm} $. This may be important for the assumptions on the error model, as will be explained in the discussion section.
Measured dispersion relations $ v_{ph}(f) $ for the two block waveguides are shown in \cref{fig:dispersion_wvg01,fig:dispersion_wvg02}. For both cases, applying a least-squares routine with \cref{eq:metallicWvgDisp} and the radius as free parameter led to reasonable fits and effective radii of $ \SI{498}{\mu m} $ and $ \SI{495}{\mu m} $, respectively. The dispersion unravels the actual geometry inside, in contrast to the radii of $ \SI{511 \pm 10}{\mu m} $ and $ \SI{505 \pm 7}{\mu m} $ measured at the apertures. While the measurements do not match within the uncertainty range around the prediction, the almost identical shape indicates a systematic error leading to the offset between prediction and measurement. Due to the drilling artifacts on the channel opening observed on the microscope images, the systematic error is attributed to the measurement of the radii. It is reasonable to assume that the channel widens close to the opening as the drill head was less constraint in space.
Both structures were measured with the blunt needle as obstacle as well as an $ \SI{0.80}{mm} $ blue steel wire. Deviations in $ v_{ph}(f) $ from different obstacles were at maximum $ \SI{0.3}{\percent} $. 

\begin{figure}
	\includegraphics[width=0.43\textwidth]{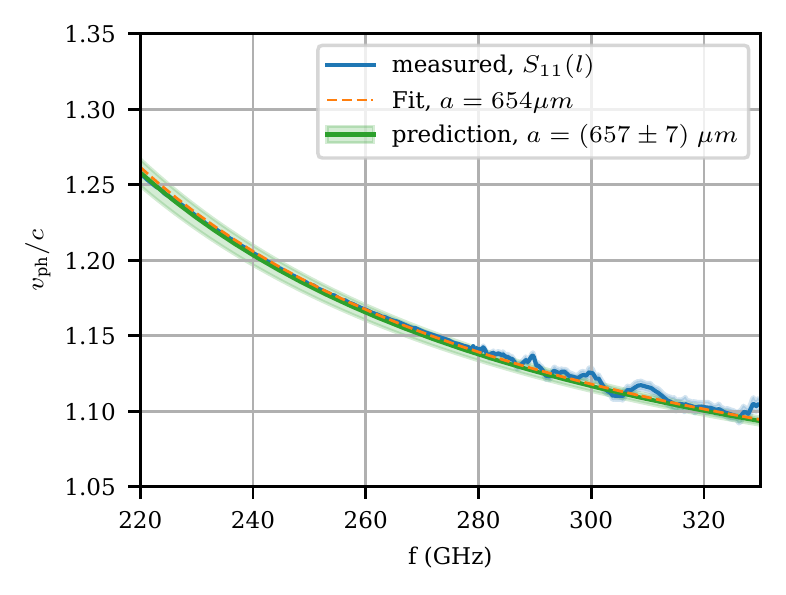}		
	\caption{Measurement of split-block waveguide analyzed by the four term error model, \cref{eq:iThs11fromNtwk}, and prediction.}
	\label{fig:dispersion_wvg04}
\end{figure}

In contrast, the measured dispersion $ v_{ph} $ of the split block waveguide, \cref{fig:dispersion_wvg04}, matches very well the prediction based on $ \SI{657 \pm 7}{\micro \meter} $ for frequencies below $ \SI{280}{GHz} $. The fit of the measurement with the analytical model, \cref{eq:metallicWvgDisp}, results in a radius of $ a = \SI{654}{\micro \meter} $ which is well within the uncertainty range of the radius measured by microscopy. The standard deviation of the fit error is below $ \SI{0.1}{\micro \meter} $ which shows how well the measurement and model match. However, above $ \SI{280}{GHz} $ the measured phase velocity clearly deviates from the fit, as well as from the prediction. The reason for that became clear when looking at the scan data for a single frequency point, for instance at $ \SI{283}{GHz} $. Although not shown here explicitly, $ S_{11}(l) $ clearly follows a beat pattern, indicating a second mode being present. When applying a discrete Fourier transform, two peaks of almost identical amplitude were identified in the reciprocal space. This explains why the model of \cref{eq:iThs11fromNtwk} fails, as it assumes a single mode propagation in the waveguide. The discrete Fourier transform is insufficient in resolving a smooth change in $ \beta (f) $ since the reciprocal space is sampled in discrete points of $ 1/(N \Delta l) $, where $ N $ refers to the number of obstacle positions and $ \Delta l $ is the step size. Zero-padding shifts the spectral sampling points, but does not improve the resolution and is still limited by the truncation artifacts. \gls{fdm}~\cite{mandelshtam_harmonic_1997, mandelshtam_erratum_1998} is not limited to discrete steps and therefore used to study the multi-mode behavior of the waveguide, using the \textit{harminv}~\cite{steven_g_johnson_nanocompharminv_2020} implementation. For interested readers, section 2 of Ref. \cite{mandelshtam_fdm_2001} provides a more rigorous comparison of the discrete Fourier transform and the \gls{fdm}.
As the \gls{fdm} is modeling the input signal as a sum of sinusoids, it is not able to properly account for the errors introduced in Sec. \ref{sec:fourTermErrorModel}.
But as seen before in \cref{fig:simulated_raw_s11}, the $ 1/(2 \beta)$ periodicity in $ S_{11}(l) $ is at least similar to a sinusoid if the error terms $ b $ and $ c $ are small.

\begin{figure}
	\includegraphics[width=0.43\textwidth]{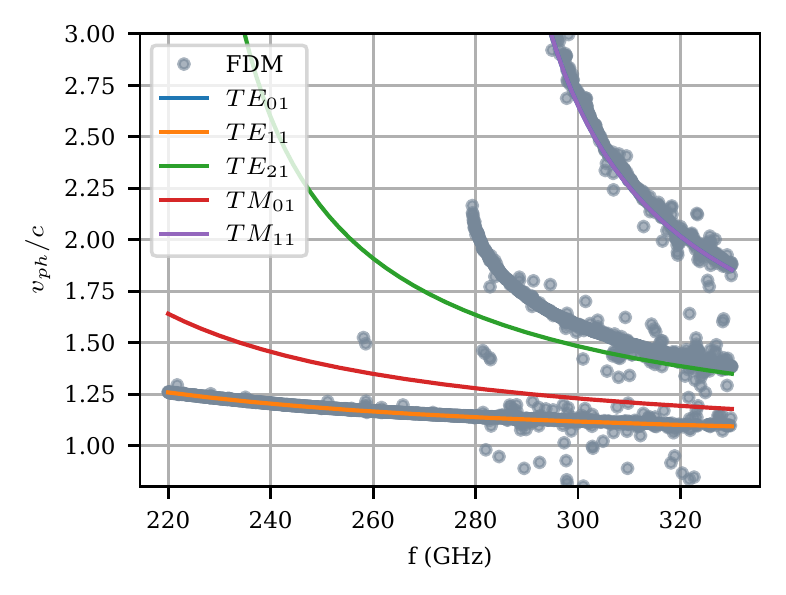}
	\caption{Multimode analysis based on the filter diagonalization method, together with expected analytical higher order mode dispersions. $TE_{01}$ and $ TM_{11} $ have identical dispersion relations.}
	\label{fig:multimode_disp_wvg04}
\end{figure}

\cref{fig:multimode_disp_wvg04} compares the modes found by \gls{fdm} with analytical expected dispersion relations of the next higher order modes. First, the \gls{fdm} finds several modes which are not following any dispersion line and seem to be spurious. This may be due to the sinusoids not properly representing the error model. Second, the majority of modes follow dispersion lines with respect to frequency. As expected, the fundamental $ TE_{11} $ mode dominates and is present over the entire frequency range. Above $ \SI{280}{GHz} $, a second mode is present which ranges from $ v_{ph} \approx \SI{2.2}{c} $ to $ \approx \SI{1.4}{c} $. Even a third dispersion line is observed above $ \SI{295}{GHz}$. Finally, the second dispersion line does not match with an analytical mode. In contrast to the first and the third line, which coincide very well with $ TE_{11} $ and $ TE_{01} $ modes, the discrepancy between the second line and the expected $ TE_{21} $ mode has not yet been understood.


\subsection{Characterization of unknown waveguides}
\label{sec:unknownWvgExperiments}
While the previous section compared measured dispersions of known \gls{dut}s to the analytical predictions, the method was also applied to devices with unknown dispersion. Regarding phase synchronous mode propagation with particles, purely metallic waveguides are unsuitable as $ v_{ph} > c $. By introducing a dielectric lining the modes mainly become hybrid, but with reduced phase velocity. With a given set of the dielectric permittivity $ \varepsilon_{r} $, the inner hollow core radius $ a $, and outer radius $ b $ of the metallic wall, the dispersion can also be computed analytically~\cite{zou_x-band_2001}. However, this requires a precise knowledge of the three parameters which is often not feasible, and a direct measurement of the dispersion is necessary.

\begin{figure}
	\includegraphics[width=0.43\textwidth]{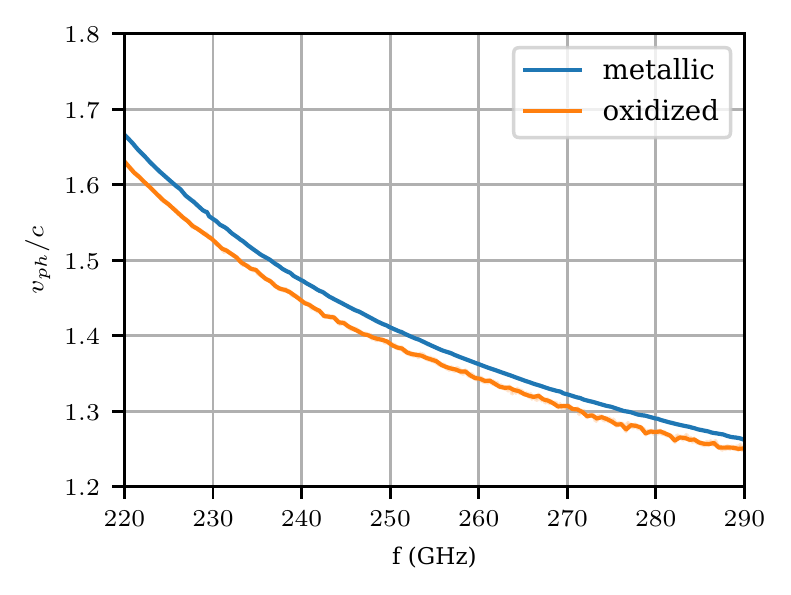}
	\caption{Comparing copper block waveguide with additional oxide layer. The frequency range is truncated to $ \SI{290}{GHz} $ due to very low signal above this threshold, as seen already partially by the ripples. }
	\label{fig:dispersion_wvg09_w_w01}
\end{figure}

\begin{figure}
	\centering
	\includegraphics{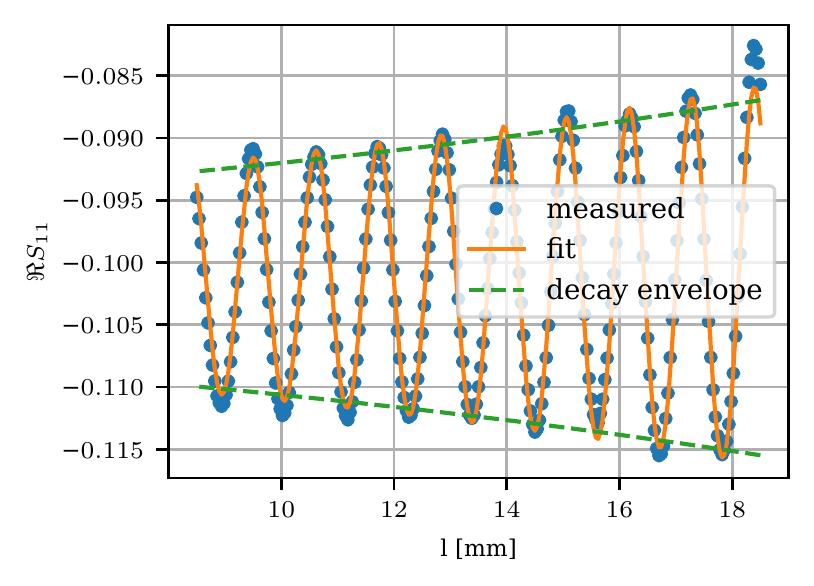}
	\caption{Real part of the measured reflection coefficient $ S_{11} $ depending on the obstacle position of the oxidized waveguide. The fit model is extended with a non-vanishing attenuation coefficient $ \alpha $. The envelope is based on $ \beta=0 $. Note that the generator is to the right side of the plot which is why the attenuation appears as an exponential growth.}
	\label{fig:raw_data_attenuation_wvg09}
\end{figure}

As a first unknown device, the copper block waveguide from \cref{fig:dispersion_wvg01} was oxidized on the inside by browning (Ballistol Nerofor). The thickness of the oxide layer is unknown as well as its permittivity in the frequency range of interest. The measurement was conducted with the same settings as before oxidization. \cref{fig:dispersion_wvg09_w_w01} shows the dispersion in comparison to the one of the purely metallic waveguide. As expected, the phase velocity has decreased. The drop in $ v_{ph} $ is about $ 0.04c $ at $ \SI{220}{GHz} $ and decreases towards higher frequencies, i.e. about $ 0.01c $ at $ \SI{290}{GHz} $. Here, the measurement above $ \SI{290}{GHz} $ was clipped. Significant loss above this threshold led to a poor fit of the model in \cref{eq:s11fromNtwk}. This loss was already present at low frequencies, while being small enough to allow for a good fit. In \cref{fig:raw_data_attenuation_wvg09}
the real part of $ S_{11}(l_{obs}) $ at $ \SI{220}{GHz} $ is shown where the exponential decay is observed. Note that, here, the position $ l_{obs} $ refers to the distance the obstacle has been moved in from the backside, which is why the attenuation is represented as an exponential growth. The fit of the four term error model includes explicitly a non-zero attenuation coefficient, $ \beta = \SI{2.828 \pm 0.0009}{mm^{-1}}, \alpha = \SI{0.025 \pm 0.0009}{mm^{-1}} $. The envelope does not properly cover the maxima of the fit since setting $ \beta = 0 $ does not account for the increase and shift of the amplitude due to the error terms $ b $ and $ c $. But the envelope follows very well the trend of the measured data. In addition to attenuation, at high frequencies the error model also fails due to higher order modes. 

\begin{figure}
	\includegraphics[width=0.43\textwidth]{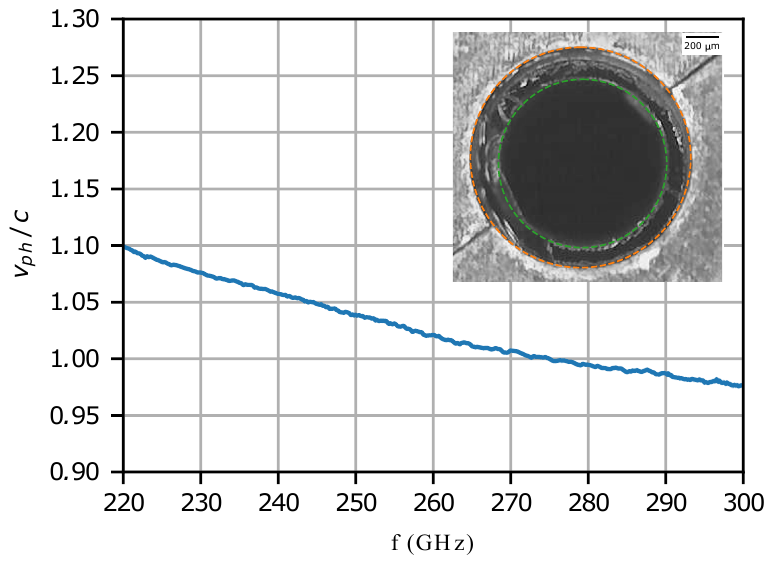}
		\caption{Phase velocity dispersion of a dielectric loaded waveguide using a 3D printed capillary. Inlet: cross sectional view of the waveguide aperture.}
	\label{fig:disp_diel_wvg13}
\end{figure}

While the oxide layer decreased the phase velocity, it is not sufficient to reach the phase synchronous case of $ v_{ph} \leq c $. A dielectric loading was created in the split-block waveguide by inserting a 3D-printed polymer capillary. The relative permittivity of the resin used for printing (Moiin Tech clear) was measured in advance~\cite{kellermeier_towards_2020}, being $ \varepsilon_r \approx 2.78 - i 0.07 $ in the frequency range of interest. The wall thickness was estimated as $ \SI{150}{\micro \meter} $. Both values are too inprecise in order to predict the dispersion relation, and the printing process was not fully optimized such that support residuals on the capillary distort the ideal shape, see the inlet of \cref{fig:disp_diel_wvg13}.

The wire with a diameter of $ \SI{0.5}{mm} $ was used as an obstacle and scanned over a $ \SI{10}{mm} $ range inside the waveguide. The resulting phase velocity dispersion is shown in \cref{fig:disp_diel_wvg13}. $ v_{ph} $ ranges between $ \SI{1.10}{c} $ at $ \SI{220}{GHz} $ and $ \SI{0.98}{c} $ at $ \SI{300}{GHz} $ and crosses $ v_{ph} = c $ at $ \SI{275}{GHz} $.


\subsection{Uncertainty analysis}
Although not visible on the chosen axis scale, dispersion lines presented in the previous section include an uncertainty band around $ v_{ph}(f) $ in the plots. This demonstrates already the level of accuracy of the measurements.

To assess the uncertainties, each frequency sweep was repeated $ 10 $ times and the sample mean $ S_{11} $ was gathered together with a $ \SI{99}{\percent} $ confidence interval. Due to the low sampling number, the sample standard error was weighted with the proper Student's t of $ 3.25 $. The absolute error in $ |S_{11}| $ did not exceed $ \num{1e-3} $ across the entire frequency band of interest, while the phase error was about $ \SI{0.03}{\degree} $.

To assess the uncertainty for $ \beta $ and $ v_{ph} $ in an obstacle scan, the dataset $ S_{11}(l) $ with its uncertainty was Monte Carlo sampled, assuming normal distribution. Each randomized copy of the dataset was applied to the fit routine of \cref{eq:iThs11fromNtwk} to extract $ \beta $ and its standard deviation. The sampling and fitting was repeated for each frequency point of $ S_{11}(l, f) $. For all measurements presented, relative uncertainties in the phase velocity are below $ \SI{0.05}{\percent} $. However, they are not shown in the plots. The standard deviation in the parameter estimate of the fit was about an order of magnitude larger. This error is not a figure of merit for the measurement uncertainty, but a measure of how well the model describes the dataset. Systematic errors due to the limits of the four-term error model dominate over statistical uncertainties.

In cases in which the mismatch of the model and the data was more severe, the fit error was substantially larger. In \cref{fig:dispersion_wvg04}, above $ \SI{280}{GHz} $ the error band enveloping the dispersion line is clearly visible and covers a range of $ \Delta v_{ph} \approx 0.02c$. The following section discusses the limits of the experimental setup and the model.

\section{Discussion}
Here, the limitations of the underlying model are discussed first. Potential improvements in the experimental design and the \gls{dut}s are considered in the subsequent section.

\subsection{Limits of the error model}
The most limiting assumption in the network model of \cref{fig:errorModelNtwk} is the obstacle being a single port device. Modes propagating further downstream of the obstacle are supposed to be absorbed or radiated away at the end of the transmission line, such that no reflections behind the obstacle interfer with the reflections from the $ Q $-network. An ideal obstacle for which the assumption is always true, would be a short circuit. No power would propagate behind the obstacle. Such devices exist in phase shifters for standard waveguides, e.g. WR-3.4. But for non-standard arbitrary shape waveguides this is technically challenging as the short device has to be non-contacting and movable.

Considering propagation behind the obstacle, the wire and the wall form another transmission line of different wave impedance and propagation constant. The transition at the waveguide exit causes another reflection and is considered as a one-port device. In the network model, $ Q $ extends to a two-port device, including the impedance mismatch between the two lines. In total, this makes $ 5 $ additional unknown parameters which would have to be included in the measured $ S_{11}^{(m)} $ of \cref{eq:s11fromNtwk}. The extended nine-term error model is a better representation of the experimental setup, but infeasible to solve. The least squares algorithm would have to optimize 16 real parameters, as the error terms are complex numbers.

To suppress mode propagation downstream, the diameter of the obstacle should be as large as possible without scratching the wall. In \cref{fig:disp_diel_wvg13} the dispersion line is not as smooth as in the other measurement, which is attributed to interference with downstream reflections. Here, the wire is only $ \SI{0.5}{mm} $ thick which is only half the size of the hollow core. In the other measurements the remaining air gap between obstacle and walls was half the size, about $ \approx \SI{100}{\micro \meter} $.

Another limiting aspect has been shown in \cref{fig:dispersion_wvg04,fig:multimode_disp_wvg04}, the excitation of higher order modes. The beat pattern was observed in several obstacle scans, but often small enough to be negligible, e.g. as in \cref{fig:raw_data_attenuation_wvg09}. The four-term error model is designed for single mode operation. Small higher order mode contributions in the reflection coefficient lead to systematic errors in the fit. An extension requires a summation over all modes excited by the current frequency, including error terms $ a_{i}, b_{i}, c_{i} $ and wavenumber $ \beta_{i} $ where $ i=0,1,2, \dots $ refers to the mode order. Since all the terms are unknown the sum cannot be fitted anymore in general. For the special case of only two modes, one could approach the problem by fitting the fundamental mode first, subtracting the fit from the data, and apply the model again to the remaining signal for the second mode. Subtracting the fit result from the original data and fitting the remaining set again refines the parameter estimates for the fundamental mode. By iterating between the two modes, the error terms could be determined more accurately. 

However, extending this scheme for more than two modes is not straight-forward. Assuming a sinusoidal shape of $ S_{11}(l) $, as it was done with applying the \gls{fdm}, is useful for finding a first estimate. It is sufficient to compare the dispersions of different modes. But as it does not properly include the asymmetric distortion due to the terms $ b $ and $ c $, it may fail in precision. Comparing the analysis presented in \cref{fig:dispersion_wvg04} and \cref{fig:multimode_disp_wvg04} up to a frequency of $ \SI{250}{GHz} $, they deviate by only $ \SI{0.08}{\percent} $ on average. But at few selected points the deviation is on the order of $ \sim \SI{1}{\percent} $, showing the discrepancy in accuracy between the error model analysis and the \gls{fdm} analysis.
 
Transitions also introduce systematical errors which is why the obstacle position is supposed to be scanned at least three wavelengths away from the coupler and the waveguide exit. As a test, the capillary loaded waveguide was studied with a scan covering partially the metallic-only section, the transition, and the dielectric loaded section. The phase velocity was determined with different truncations of the scan $ S_{11}(l) $ relative to transition, one directly starting from the transition, one being $ \SI{2}{mm} $ away, and one being $ \SI{3}{mm} $ away from the metallic section. $ v_{ph} $ differed about $ 0.01c $ between each case. With further increase in distance no significant change was observed. The found transition region is only valid for the certain case of step-transition and cannot be generalized to arbitrary transitions. But it gives already a zeroth order estimate for the impact on the measurement.

\subsection{Experimental and device limits}
Besides the methodological limitations, the setup and the structures under test are not ideal. The setup is not optimized for maximum coupling. Usually, horn-couplers are often designed for ideal plane wave coupling, i.e. maximum aperture efficiency as considered for the optimum horn condition~\cite{balanis_conical_2012, king_radiation_1950}. 
Further improvement in coupling is achieved by extending it to Gaussian beam optics~\cite{murphy_aperture_1988}. However, there is no universal condition. The ideal horn design depends on the incident Gaussian beam, or the quasi-optical layout has to match the given horn design.

Due to the lack of Gaussian optics in the current setup the first test devices, \cref{fig:allWvg_imgs}(a) and \cref{fig:allWvg_imgs}(b), were based on the optimal horn condition. The third one is manufactured with a different horn geometry to match the optical layout of another setup. In both cases, the horn not only captures the incident power, but also causes a phase shift. The effect was simulated with CST Microwave Studio for an exemplary case. A waveguide was excited by an incident Gaussian beam, once with and once without horn antenna. The phase at the output port was compared for both cases, showing 
a difference of $ \approx \SI{110}{\degree} $, almost constant over the excited frequency band from $ \SI{265}{GHz} $ to $ \SI{275}{GHz} $. For a $ \SI{20}{mm} $ long waveguide this leads to an error of $ \approx \SI{1.5}{\percent} $ in the phase velocity. The current scanning reflection method overcomes this systematic error by attributing the horn-related phase shift to the error terms.

With the aim for phase synchronous mode propagation and efficient THz-particle interaction in an accelerator based application, the dielectric lining requires significant manufacturing improvements. On the one hand, mechanical residues from 3D-printing  have to be reduced. The circular shape of the hollow core is significantly distorted along the longitudinal axis, causing scattering of the propagating wave. On the other hand, the polymer is lossy and requires further treatment to reduce absorption. Different materials such as fused silica, are more desirable, also because of their higher rf-breakdown limits. But they also increase the machining complexity for an integrated coupling device.

\section{Outlook}
Due to coupling to the fundamental mode, the current setup characterizes modes suitable for transverse deflection of bunches~\cite{lemery_transverse_2017}. For characterizing the accelerating $ TM_{01} $ mode the setup will be equipped with a segmented half-wave plate to match the incident field pattern to the waveguide pattern. PTFE lenses are added for focusing the beam into the coupler. However, both optical elements reduce the spanned frequency bandwidth due to being rather monochromatic.

Future structures to be studied include another polymer based dielectric lined waveguide, but instead of being inserted into a metallic structure, the combined polymer horn capillary will be coated with a metallic layer. In addition, fused silica capillaries will be coated and inserted in metallic horn structures.

Potentially more complex devices, such as a tapered dielectric waveguide for synchronous acceleration of low-energy particles \cite{lemery_synchronous_2018}, or a sectioned waveguide for longitudinal phase space synthesis~\cite{mayet_longitudinal_2020}, could be explored with the help of windowing the obstacle scan, resulting in a spectrogram representing the change of $ \beta $ as it varies with position.

\section{Conclusion}
A new error model is developed with the purpose of self-calibrating reflection measurements from monolithic integrated hollow waveguides via free-space coupling. Reflections from unmatched couplers, from surrounding mounting elements as well as free-space losses and phase jumps at the reflecting movable obstacle are included in the error-terms. The model is applied to a simulation to investigate the deviations from the input error-terms and phase velocity.

Experiments with metallic waveguides of known cross section are in good agreement with analytical predictions. While the measurements provide a precision on the level of $ \approx \SI{0.05}{\percent} $, systematic errors not covered by the error model currently limit the accuracy to $ \approx \SI{0.5}{\percent} $. One of the systematic errors is introduced by reflections downstream of the obstacle. Comparison measurements between a cannula and a blue steel wire as obstacles deviate up to $ < \SI{0.3}{\percent} $. A dedicated choke-type piston as movable short may reduce the downstream reflections in future. In addition, the error model breaks down in case of higher order mode excitation.

The method was also applied to dielectric-lined metallic waveguides with both a copper oxide and a polymer lining, yielding reduced phase velocities. In the case of the thicker 3d-printed polymer, we measured a phase velocity equal to the speed of light at a frequency of $ \sim \SI{275}{GHz} $, compatible for efficient beam acceleration and manipulation with relativistic beams.

\begin{acknowledgments}
	The authors are grateful for the helpful scientific discussions with Markus Hünning and Thomas Vinatier, the technical inputs from Frank Mayet and Olaf Rasmussen, and experimental discussions with Hannes Dinter, Christian Henkel and Immo Bahns.
	
	This work has been supported by the European Research Council under the European Union’s Seventh Framework Programme (FP/2007-2013)/ERC Grant Agreement n. 609920. 
	
	All figures and pictures by the author(s) are published under a \href{https://creativecommons.org/licenses/by/4.0/}{CC-BY} license.
\end{acknowledgments}
\section*{Supplemental Material}
For reproducibility a python-based implementation of the 4-term error model used for the analysis is attached to the paper, as well as the raw data sets of the presented experimental results.

\appendix
\section{Derivation of the error model}
The waves between the networks in the flow graph \cref{fig:errorModelNtwk} are denoted as $ (a_1, b_1) $ before the $ P $-network, $ (a_2, b_2) $ between $ P $- and $ S $-network, and $ (a_3, b_3) $ between $ S $- and $ Q $-network. For each interconnection the reflection coefficient from the right-sided device is denoted as $ \rho_i = \sfrac{b_i}{a_i} $. When going from right to left in the network, each subnetwork represents a Möbius transform of the reflection coefficient. Applying the transform from interconnection $ 3 $ to $ 2 $,
\begin{equation}
	\rho_3 = \dfrac{b_3}{a_3} = Q_{11}, \quad 
	\left[\begin{array}{l}
		b_{2} \\
		a_{3}
	\end{array}\right]=\left[\begin{array}{ll}
		0 & k\\
		k & 0
	\end{array}\right]\left[\begin{array}{l}
		a_{2} \\
		b_{3}
	\end{array}\right]
\end{equation} 
\begin{equation}
	\rho_2 = \dfrac{b_2}{a_2}=k^2 Q_{11}.
\end{equation}
The transform from interconnection $ 2 $ to $ 1 $ results in
\begin{align}
	\left(\begin{array}{l}
		b_{1} \\
		a_{2}
	\end{array}\right)=\left[\begin{array}{ll}
		P_{11} & P_{12} \\
		P_{21} & P_{22}
	\end{array}\right] \left(\begin{array}{l}
		a_{1} \\
		b_{2}
	\end{array}\right) 
	\quad \Rightarrow
	\rho_1 = \dfrac{P_{11} a_1 + P_{12}b_2}{a_1} \\
	\rho_1 = P_{11} + \rho_{2} \dfrac{P_{21} P_{12}}{(1- P_{22} \rho_2)} = P_{11} + Q_{11} \dfrac{P_{21} P_{12}}{k^{-2} - P_{22} Q_{11}}
\end{align}
%

\end{document}